\documentclass[twocolumn,showpacs,prb]{revtex4}

\usepackage{graphicx}
\usepackage{amsmath}
\usepackage{amssymb}
\usepackage{dsfont}
\usepackage{braket}

\DeclareMathOperator{\e}{e}
\newcommand{\Jqt}{J_\mathrm{q}^{-1}}
\newcommand{\Jq}{J_\mathrm{q}}
\newcommand{\dt}{\mathrm{d}t}

\usepackage{hyperref}
\hypersetup{%
  breaklinks = {true}, 
  colorlinks = {true}, 
  citecolor = {blue},  
  filecolor = {black}, 
  linkcolor = {red},   
  urlcolor  = {blue},  
  pdfauthor = {\textcopyright\ Daniel Stanek, Carsten Raas, and G\"otz S. Uhrig},
  pdfstartpage = {1},
  pdfstartview = {FitH}, 
}

\begin{document}

\title{From quantum-mechanical to classical dynamics in the central-spin model}

\author{Daniel Stanek}
\affiliation{Lehrstuhl f\"{u}r Theoretische Physik I, Technische Universit\"at Dortmund, 44221 Dortmund, Germany}
\author{Carsten Raas}
\affiliation{Lehrstuhl f\"{u}r Theoretische Physik I, Technische Universit\"at Dortmund, 44221 Dortmund, Germany}
\author{G\"otz S. Uhrig}
\email{goetz.uhrig@tu-dortmund.de}
\affiliation{Lehrstuhl f\"{u}r Theoretische Physik I, Technische Universit\"at Dortmund, 44221 Dortmund, Germany}

\begin{abstract}
  We discuss the semiclassical and classical character of the dynamics of a
  single spin $1/2$ coupled to a bath of non-interacting spins $1/2$. On the
  semiclassical level, we extend our previous approach presented in D. Stanek,
  C. Raas, and G. S. Uhrig, Phys.\ Rev.\ B \textbf{88}, 155305 (2013) by the
  explicit consideration of the conservation of the total spin. On the
  classical level, we compare the results of the classical equations of
  motions in absence and presence of an external field to the full quantum
  result obtained by density-matrix renormalization (DMRG). We show that for
  large bath sizes and not too low magnetic field the classical dynamics,
  averaged over Gaussian distributed initial spin vectors, agrees
  quantitatively with the quantum-mechanical one. This observation paves the
  way for an efficient approach for certain parameter regimes.
\end{abstract}


\maketitle

\section{Introduction \& Motivation}

Single electron or hole spins confined in a quantum dot are promising
candidates for the realization of quantum bits
(qubits).\cite{Loss1998,hanso07,urbas13} The requirements for a system to be a
good candidate for a quantum computer are summarized in the well-known
criteria of DiVincenzo.\cite{DiVinc2000} One essential ingredient is the
existence of long decoherence times to dispose of long storage times for the
quantum information or to make a large number of gate operations
possible. Hence, a detailed understanding of the real-time evolution of the
qubit induced by the interaction with its environment is crucial.

In bulk solids, the dominating contribution to the decoherence of an
electronic spin, taken as qubit, is usually based on spin-orbit coupling. However,
it has been shown that the relaxation \cite{Khaets2000,Khaets2001} and the
dephasing\cite{Golova2004} due to spin-orbit coupling of an electron spin
confined in a quantum dot is strongly suppressed. Instead, the hyperfine
coupling between the electron spin and the surrounding nuclear spins in the
dot is the major player.\cite{Merkul2002,Schlie2003} This situation is well
captured by the central-spin or Gaudin model (CSM)\cite{Gaudin1976,Gaudin1983}
\begin{align}
  H&=-h_0S^z_0+\vec{S}_0\cdot\sum\limits_{i=1}^N J_i\vec{S}_i,
  \label{eq:02_Gaudin}
\end{align}
where a central spin $\vec{S}_0$ in an external field $h_0$ interacts with $N$
non-interacting bath spins $\vec{S}_i$. The coupling constants $J_i$ are
distributed randomly because their values are proportional to the probability
$|\psi(\vec{r}_i)|^2$ [$\psi(\vec{r})$ being the wave vector of the confined
electron] that the electron is present at the site of the nucleus $i$ located
randomly in the dot.\cite{Merkul2002,Schlie2003} It is convenient to represent
the bath by the operator
\begin{subequations}\label{eq:02_Gaudin_A}
  \begin{align}
    \vec{A}&:=\sum\limits_{i=1}^N J_i\vec{S}_i,\label{eq:02_A}
  \end{align}
  which acts on all bath spins weighted with their corresponding coupling
  constant. It can be interpreted as an effective three-dimensional magnetic
  field created by the bath spins called the \textsc{Overhauser field}. The
  Hamiltonian \eqref{eq:02_Gaudin} simplifies to
  \begin{align}
    H&=-h_0^{\phantom{z}}S^z_0+ \vec{S}_0\cdot\vec{A}.
  \end{align}
\end{subequations}
	
We take the external field $h_0$ to be restricted to the central spin
$\vec{S}_0$ because the Zeeman splitting of the nuclear spins is much smaller
due to the small magnetic moments of the nuclei. The magnetic moment of the
electron is about three orders of magnitude larger. The dipolar interaction
between nuclear bath spins is not considered because it affects the physics of
the model only on a much larger time scale than the hyperfine
interaction.\cite{Merkul2002} The intrinsic energy scale of a typical
self-assembled quantum dot is of the order of
$10^{-5}\,\mathrm{eV}$.\cite{Merkul2002,Schlie2003,lee05,petro08} Experiments
are usually performed at temperatures in the range $T=6$--$50\,\mathrm{K}$
corresponding to thermal energies $k_\mathrm{B}T\approx
10^{-4}$--$10^{-3}\,\mathrm{eV}$.\cite{Hernan2008,Greili2009} Thus, the energy
scale of the thermal fluctuations is at least one order of magnitude larger
than the intrinsic energy scale of the spins in the quantum dot. This implies
that the temperatures occurring in experiment can be taken to be infinite so
that the nuclear spins are initially completely unpolarized.

The real-time evolution of the central spin can be computed by numerical
techniques such as exact diagonalization\cite{Schlie2003,Cywins2010} or the
Chebyshev expansion.\cite{Dobrov2003,Hackma2014} Analytical solutions can be
derived by means of the Bethe ansatz for strongly polarized initial
states.\cite{Bortz2007a,Bortz2007b} Recently, a combination of the algebraic
Bethe ansatz and Monte Carlo sampling was able to access unpolarized initial
mixtures with up to 48 bath spins.\cite{Fariba2013,Fariba2013a} Much larger
systems can be treated by DMRG, but generically the reliability of the results
is limited in time due to the growing truncation error occurring in the DMRG
sweeps.\cite{Stanek2013} We will see an exception to this rule for strong
magnetic fields where this restriction turns out to be much less severe.
Approaches based on a non-Markovian master equation give access to larger bath
sizes, but they are limited to sufficiently strong external
fields.\cite{Coish2004,Breuer2004,Fische2007,Ferrar2008,Barnes2012} With
cluster expansion techniques, the closely related problem of the dephasing of
the central spin due to spectral diffusion can be
treated.\cite{Witzel2005,Witzel2006,Maze2008,Yang2008a,Yang2009,Witzel2012}

Moreover, the dynamics of the central spin was frequently studied on the level
of semiclassical or classical models. These methods comprise a variety of
approaches, for instance, the replacement of the bath by an effective
time-dependent field.\cite{Erling2002,Erling2004,Merkul2002,Stanek2013} As a
first approximation, the bath may be regarded as frozen, i.e, the Overhauser
field is constant. Subsequently, random fluctuations of the bath due to the
interaction with the central spin can be included.\cite{Merkul2002} Assuming
that the Overhauser field can be described as stochastic field, the
fluctuations of the central spin can be found from solution of the Bloch
equation of the Langevin type.\cite{Glazov2012,Hackma2014a}

Furthermore, it was argued that the classical trajectories of the CSM
resulting from the saddle point approximation of the spin-coherent path
integral representation describe the central-spin dynamics well because the
quantum fluctuations become less important for large numbers of bath
spins.\cite{Chen2007} Similarly, a large number of bath spins was accessed by
combining the so-called $P$ representation of the density matrix with
time-dependent mean-field theory.\cite{AlHass2006,Zhang2006}

The goal of the present paper is to study to what extent the quantum-mechanical dynamics of the central spin can be described by simpler models. We
will focus on the semiclassical and the classical model. By comparison of
fully quantum-mechanical results with the results from these two models we
establish for which regimes the simpler calculations can be regarded as
essentially equivalent to the quantum-mechanical ones. Thereby, certain future
investigations can be done much more efficiently because the evaluation of the
simpler models is sufficient. To our knowledge, a detailed comparison between
the classical and quantum behavior in the CSM has not been conducted so far.

For our proof-of-principle study, we restrict ourselves to spins $S=1/2$ and
discuss a generic uniform distribution $J_i\in[0,J_\mathrm{c}]$ where the
cutoff $J_\mathrm{c}$ is determined by the total energy scale $J_\mathrm{q}$
of the CSM. Since we deal with an unpolarized bath, the energy scale is given
by the root of the second moment of the couplings \cite{Merkul2002}
\begin{align}
  J_\mathrm{q}^2&:=\sum\limits_{i=1}^N J_i^2.
  \label{eq:02_Jq}
\end{align}
Consequently, the natural unit of time for the fast dynamics is given by
$1/J_\mathrm{q}$ which we use in the sequel (we set $\hbar$ to unity). We pick
equidistant couplings from the interval $[0,J_\mathrm{c}]$
\begin{align}
  J_i&=\sqrt{\frac{6N}{2N^2+3N+1}}\frac{N+1-i}{N} J_\mathrm{q},
  \label{eq:02_Ji}
\end{align}
where $ i\in\{1,\ldots,N\}$ to represent a uniform distribution of the $J_i$
in $[0,J_\mathrm{c}]$.

The paper is organized as follows: In Sect.~\ref{sec:semi}, we discuss the
semiclassical approach. The described ansatz incorporates the conservation of
the total spin leading to a good description up to intermediate time scales.
The classical equations of motion are discussed in Sect.~\ref{sec:classical}
with and without an external magnetic field. A detailed comparison with the
DMRG results for the quantum model is presented to verify the validity of the
classical picture. Finally, our findings are concluded in
Sect.~\ref{sec:conclusion}.

\section{Semiclassical model}
\label{sec:semi}

The classical character of the bath is to be expected for a large number of
bath spins. This has been made rigorous by an analytic argument for the
operator norm and the commutator of the Overhauser field.\cite{Stanek2013}
Further support stems from the comparison to DMRG results for zero external
field. Hence one may replace the operator $\vec{A}$ of the Overhauser field
by a classical field $\vec{\eta}(t)$ to obtain the semiclassical Hamiltonian
\begin{align}
  H_\mathrm{sc}&=\vec{\eta}\left(t\right)\cdot \vec{S}_0.
  \label{eq:04_HSC}
\end{align}
The central spin $\vec{S}_0$ is still treated on the quantum level. In
addition, the dynamics of the classical field $\vec{\eta}$ is assumed to be
stochastic and of Gaussian statistics according to the central limit theorem.
We stress that this assumption neglects any backaction effects of the central
spin on the fluctuations of the Overhauser field. The fluctuations are fully
defined by the autocorrelation functions
\begin{subequations}
  \begin{align}
    g_{\alpha\beta}\left(t_1-t_2\right)&=\overline{
      \eta_\alpha\left(t_1\right)\eta_\beta\left(t_2\right)}
  \end{align}
  with $\alpha, \beta\in\{x, y, z\}$ and their mean values
  \begin{align}
    \overline{\eta_\alpha\left(t\right)}&=0 .
  \end{align}
\end{subequations}
Without loss of generality, the mean values are taken to vanish because a
finite value can be interpreted as a contribution of an external magnetic field.
In passing from the quantum to the semiclassical model, the autocorrelation
functions $g_{\alpha\beta}(t)$ are identified with the autocorrelations
$\braket{A^\alpha(t)A^\beta(0)}$ of the Overhauser field, see
Ref.~\onlinecite{Stanek2013} for details.

In previous works, the comparison between the semiclassical model and the
quantum model has revealed a very good agreement on short time scales up to
$t\approx10\,\Jqt$ or for finite frequencies.\cite{Stanek2013,Hackma2014a}
However, for longer times, the autocorrelation function of the central spin
always displays a pronounced decay in the semiclassical calculation which does
not coincide with what is found in quantum-mechanical calculations, for an
example see the red dashed curves labeled ``Langevin 1'' in Fig.~\ref{fig:2}. The neglect of conservation laws of the full quantum model due to
the semiclassical treatment is one reason for this behavior as we will
illustrate below.

For details of the calculation ``Langevin 1'', we refer the reader to Ref.~\onlinecite{Stanek2013} and to Sect.~\ref{ss:zero-field} below. One key
element is that the autocorrelations of the Overhauser field are taken from
the numerical DMRG calculation.

\subsection{Conservation of the total spin}

An obvious conserved quantity in the central spin model is the total spin
\begin{align}
  \vec{I}&=\sum\limits^N_{i=0} \vec{S}_i.
  \label{eq:04_I}
\end{align}
By construction, the numerical DMRG captures the conservation of the total
spin and all other conservation laws in the CSM to the degree of its numerical
accuracy. However, this does not hold for the semiclassical model defined in
Eq.~\eqref{eq:04_HSC} which obviously does not display a conservation law for
the fluctuating Overhauser field $\vec{\eta}$.

To improve the reliability of the semiclassical results as predictions for the
quantum-mechanical calculation, we show here how the conservation of the total
spin $\vec{I}$ can be incorporated in the semiclassical model and discuss its
effect.  To this end, we study the slightly modified Hamiltonian
\begin{align}
  H'&=\vec{S}_0\cdot \sum\limits_{i=0}^NJ_i\vec{S}_i
\end{align}
for the quantum CSM. The central spin $\vec{S}_0$ has been included in the sum
which was restricted originally to the bath spins. For $S=1/2$, the additional
contribution induces only a constant shift $3J_0/4$ in the Hamiltonian. Thus,
it has no influence on the dynamics of the model. The arbitrary coupling
constant of the central spin is assigned the mean value of all couplings
\begin{align}
  J_0&:=\frac{1}{N}\sum\limits^N_{i=1} J_i.
\end{align}
Consequently, the fluctuating field
\begin{subequations}
  \begin{align}
    \vec{A}&=\sum\limits_{i=0}^NJ_i\vec{S}_i
  \end{align}
  comprises the central spin $\vec{S}_0$ in addition to the bath spins. To
  take the conservation of the total spin into account, we rewrite $\vec{A}$
  in the form
  \begin{align}
    \vec{A}&=\vec{A}_0+\Delta\vec{A}.
    \label{eq:04_A_opt}
  \end{align}
\end{subequations}
The part
\begin{subequations}
  \begin{align}
    \vec{A}_0&=J_0 \vec{I} \label{eq:04_A0}
  \end{align}
  is conserved and thus constant in time while the contribution
  \begin{align}
    \Delta\vec{A}&=\sum\limits_{i=1}^N\left(J_i-J_0\right)\vec{S}_i \label{eq:04_DeltaA}
  \end{align}
  comprises the fluctuating part.
\end{subequations}

The temporal constancy $\vec{A}_0(t)=\vec{A}_0(0)$ implies that the
correlation of the conserved part is given by the constant expression
\begin{subequations}
  \begin{align}
    \braket{A^\alpha_0\left(t\right)A^\beta_0\left(0\right)}&=J_0^2\frac{N+1}{4}\delta_{\alpha\beta}.
    \label{eq:04_g_A0}
  \end{align}
  Here we focus on the zero-field limit where all nondiagonal correlations
  vanish and the diagonal correlations are isotropic: $g(t) :=
  g_{\alpha\alpha}(t)$. However, we continue to use the general notation so that
  the present discussion can easily be extended to other symmetries.

  It is important to realize that the conserved and the fluctuating parts are
  independent at all times in the sense that their correlations vanish
  \begin{align}
    \braket{A^\alpha_0\left(t\right)\Delta A^\beta_0\left(0\right)} &=
    \braket{A^\alpha_0\left(0\right)\Delta A^\beta_0\left(0\right)} \\
    &=\frac{J_0}{4}\delta_{\alpha\beta}\sum\limits_{i=0}^N\left(J_i-J_0\right) \\
    &=0.
  \end{align}
\end{subequations}
Thus, the autocorrelation function of the Overhauser field $\vec{A}$ acquires
the form
\begin{subequations}
  \label{eq:eta_totalspin}
  \begin{align}
    g_{\alpha\beta}\left(t\right)&=J_0^2\frac{N+1}{4}\delta_{\alpha\beta} +
    \Delta g_{\alpha\beta}\left(t\right)\label{eq:04_g_opt}
  \end{align}
  with
  \begin{align}
    \Delta g_{\alpha\beta}\left(t\right)&:=\braket{\Delta
      A^\alpha\left(t\right)\Delta A^\beta\left(0\right)}.
    \label{eq:04_g_DeltaA}
  \end{align}
\end{subequations}

Next, we address the central spin $\vec{S}_0$ which is treated similarly to
the field $\vec{A}$ by splitting it into a constant and a fluctuating part
\begin{subequations}
  \begin{align}
    \vec{S}_0&=\frac{1}{N+1}\vec{I}+\Delta\vec{S}_0,\label{eq:04_S0_new}
  \end{align}
  where the fluctuating part reads
  \begin{align}
    \Delta\vec{S}_0&=\frac{N}{N+1}\vec{S}_0-\frac{1}{N+1}\sum\limits_{i=1}^N\vec{S}_i\label{eq:04_DeltaS}
  \end{align}
\end{subequations}
and the fraction $\vec{I}/(N+1)$ of the total spin $\vec{I}$~\eqref{eq:04_I}
is the conserved, constant contribution. Like for $\vec{A}$, there is no
correlation between the constant and the fluctuating parts for any time
\begin{subequations}
  \begin{align}
    \braket{I^\alpha\left(t\right)\Delta
      S_0^\beta\left(0\right)}&=\braket{I^\alpha\left(0\right)\Delta
      S_0^\beta\left(0\right)}
    \\
    &=\frac{1}{4}\frac{N}{N+1}-N\frac{1}{4}\frac{1}{N+1}
    \\
    &=0.
  \end{align}
\end{subequations}
Consequently, the autocorrelation function of the central spin $\vec{S}_0$ is
given by
\begin{subequations}
  \begin{align}
    \braket{S^\alpha_0\left(t\right)S^\beta_0\left(0\right)}&=\delta_{\alpha\beta}\frac{1}{4}\frac{1}{N+1}+
    \Delta c_{\alpha\beta}\left(t\right)
    \label{eq:confrac}
  \end{align}
  with
  \begin{align}
    \Delta c_{\alpha\beta}\left(t\right)&:=\braket{\Delta
      S^\alpha_0\left(t\right)\Delta S^\beta_0\left(0\right)}.
  \end{align}
\end{subequations}
Note that the conserved fraction [first term on the right hand side of
Eq.~\eqref{eq:confrac}] vanishes for infinitely large systems $N\to\infty$.

In the above way, the conserved part is separated from the fluctuating part
for both the central spin and the Overhauser field $\vec{A}$. We incorporate
this concept into the semiclassical model by considering the Hamiltonian
\begin{align}
  H_\mathrm{sc}'&=\vec{\eta}\left(t\right)\cdot\Delta\vec{S}_0\label{eq:04_Hsc_opt_a}
\end{align}
which refers to the fluctuating part of the central spin only. Since the
conserved part is constant, it does not enter in $H_\mathrm{sc}'$. As before,
the fluctuating field $\vec{\eta}(t)$ is a random Gaussian variable. Its
correlation function is defined by $g_{\alpha\beta}(t)$ in
Eq.~\eqref{eq:04_g_opt} involving the separate treatment of the fluctuating
and the conserved parts of $\vec{A}$ in Eq.~\eqref{eq:04_A_opt}. Thus, the
conservation of the fraction of the Overhauser field, which is proportional to
the total spin, is built in.

Inserting the expression for $\Delta\vec{S}_0$ from
Eq.~\eqref{eq:04_Hsc_opt_a}, the semiclassical Hamiltonian becomes
\begin{subequations}
  \begin{align}
    H_\mathrm{sc}'&=\sum\limits_{i=0}^N h_i
  \end{align}
  where
  \begin{align}
    h_0&:=\frac{N}{N+1}\vec{\eta}\left(t\right)\cdot\vec{S}_0
    \label{eq:04_opt_h0}\\
    h_i&:=-\frac{1}{N+1}\vec{\eta}\left(t\right)\cdot\vec{S}_i, \ \ \
    i\in\left\{1,2,\ldots,N\right\}.
    \label{eq:04_opt_hi}
  \end{align}
\end{subequations}
On this level of description, the time evolution of the bath spins $\vec{S}_i,
i>0$ is completely independent from one another. The fluctuating part $\Delta
c_{\alpha\beta}\left(t\right)$ of the autocorrelation function of the central
spin can be calculated by
\begin{subequations}
  \begin{align}
    \label{eq:weight_observ}
    \Delta c_{\alpha\beta}\left(t\right)&=\left(\frac{N}{N+1}\right)^2 \Delta
    c_{\alpha\beta}^{(0)}\left(t\right) + \frac{N}{\left(N+1\right)^2} \Delta
    c_{\alpha\beta}^{(i)}\left(t\right)
  \end{align}
  with the two independent contributions
  \begin{align}
    \Delta
    c_{\alpha\beta}^{(0)}\left(t\right)&:=\braket{S^\alpha_0\left(t\right)S^\beta_0\left(0\right)}
    \label{eq:c0} \\
    \Delta
    c_{\alpha\beta}^{(i)}\left(t\right)&:=\braket{S^\alpha_i\left(t\right)S^\beta_i\left(0\right)}
    \label{eq:ci}
  \end{align}
\end{subequations}
where the former [Eq.~\eqref{eq:c0}] acquires its dynamics from the
Hamiltonian $h_0$ in Eq.~\eqref{eq:04_opt_h0} and the latter [Eq.~\eqref{eq:ci}] acquires its dynamics from the Hamiltonian $h_i$ in Eq.~\eqref{eq:04_opt_hi}.

In total, two independent runs of the code are required for simulating this
enhanced semiclassical model which respects the conservation of the total
spin. The run for the Hamiltonian $h_0$ involves the strong coupling $\propto
N/(N+1)$ between central spin and bath. Thus, the contribution $\Delta
c_{\alpha\beta}^{(0)}\left(t\right)$ dominates the fast dynamics of the
autocorrelation function
$\braket{S^\alpha_0\left(t\right)S^\beta_0\left(0\right)}$. In contrast, the
coupling $-1/(N+1)$ between a single bath spin $\vec{S}_i$ and $\vec{\eta}(t)$
is small, in particular for large baths. Consequently, the Hamiltonian $h_i$
induces only a very slow dynamics. In addition, the weight factors in Eq.~\eqref{eq:weight_observ} imply that the fast dynamics $\propto N/(N+1)$ is
dominating anyway due to an extra factor $N$. Note, that for infinitely large
systems with $N\to\infty$ the fast dynamics is the only one remaining. The
calculations on this level of the semiclassical model are labeled ``Langevin
2'' in Fig.~\ref{fig:2}. More details of the calculations are given below in
Sect.~\ref{ss:zero-field}.

\subsection{The central spin as classical vector}
\label{ss:centrspin-classvec}

In the above introduced modification ``Langevin 2'' of the random noise
simulation, the central spin is treated on the quantum level, while the bath is
a classical variable. However, the precession of a quantum spin in an external
field is identical to the one of a classical vector in $\mathds{R}^3$ due to
the linearity of the equations of motion. It implies that the quantum-mechanical expectation values follow exactly the classical equations of motion
according to Ehrenfest's theorem. Thus, we go one step further and replace the
quantum-mechanical central spin by a classical vector in the simulations of
the semiclassical model. Otherwise, we keep the separation in a conserved
constant part and a fluctuating decaying part.

We recall the semiclassical Hamiltonian \eqref{eq:04_Hsc_opt_a} and insert the
expression for the fluctuating part $\Delta \vec{S}_0$ of the central spin
defined in Eq.~\eqref{eq:04_S0_new}
\begin{align}
  \label{eq:hamilton-langevin3}
  H_\mathrm{sc}&=\vec{\eta}\left(t\right)\left(\vec{S}_0-\frac{1}{N+1}\vec{I}\right).
\end{align}
According to Eq.~\eqref{eq:04_A_opt}, the Gaussian fluctuation $\vec{\eta}(t)$
can be written as
\begin{align}
  \label{eq:random_field}
  \vec{\eta}\left(t\right)&=J_0\vec{I}+\Delta\vec{\eta}\left(t\right).
\end{align}

Since the Hamiltonian \eqref{eq:hamilton-langevin3} is a classical
Hamiltonian, one easily deduces the two corresponding equations of motion:
\begin{subequations}
  \label{eq:04_EOM_classic}
  \begin{align}
    \frac{\mathrm{d}}{\dt}\vec{S}_0&=\vec{\eta}\left(t\right)\times\vec{S}_0
    \label{eq:04_EOM_classica}\\
    \frac{\mathrm{d}}{\dt}\vec{I}_{\phantom{0}}&=-\frac{1}{N+1}
    \vec{\eta}\left(t\right)\times\vec{I},
    \label{eq:04_EOM_classicb}
  \end{align}
\end{subequations}
where all spins are classical vectors in $\mathds{R}^3$. We also adopt the
former expression for the autocorrelation function:
\begin{subequations}
  \begin{align}
    \overline{S^\alpha_0\left(t\right)S^\beta_0\left(0\right)}&=\frac{1}{4}\frac{1}{N+1}\delta_{\alpha\beta}
    + \Delta c_{\alpha\beta}\left(t\right)
    \label{eq:04_SzSz_classic}
  \end{align}
  with
  \begin{align}
    \Delta
    c_{\alpha\beta}\left(t\right)&:=\overline{\left(S^\alpha_0-\frac{1}{N+1}
        I^\alpha\right)\Big(t\Big)\left(S^\beta_0-\frac{1}{N+1}
        I^\beta\right)\Big(0\Big)}
    \label{eq:04_DeltaC_classic}.
  \end{align}
\end{subequations}
To distinguish the latter expressions from the quantum description, the
expectation values are denoted by an overbar and not by Dirac brackets
$\left<.\right>$.

The equations of motion~\eqref{eq:04_EOM_classic} can easily be integrated
using standard methods and subroutines, for instance Runge-Kutta
integration. As in the previous section for the improved semiclassical
calculation, two independent runs of the integration are required: one for the
precession due to \eqref{eq:04_EOM_classica} for a strong coupling and another
for the precession due to \eqref{eq:04_EOM_classicb} for a weak coupling to
the random field $\vec{\eta}(t)$, Eq.~\eqref{eq:random_field}. Except for this
change, the integration can be carried out with the same code.

In the simulation, we sample the Gaussian fluctuations $\Delta\vec{\eta}(t)$
complying with the autocorrelation function $\Delta g_{\alpha\beta}(t)$
defined in Eq.~\eqref{eq:04_g_DeltaA}. However, which initial values are to be
taken for $I^\alpha$ and $\Delta S_0^\alpha(0)$? We only have information on
the averages: the mean values, corresponding to expectation values,
vanish. However, the equal-time correlations are finite and known from the
quantum-mechanical counterparts. Thus, we choose random initial values drawn
from Gaussian distributions, which reproduce for $t=0$ the quantum-mechanical
correlations. Concretely, we know the variance for a single component of the
total spin to be
\begin{subequations}
  \begin{align}
    \braket{\left(I^\alpha\left(0\right)\right)^2}&=\frac{N+1}{4}.
  \end{align}
  For a single component of the fluctuating part $\Delta S_0^\alpha(t)$, the
  initial variance at $t=0$ is given by the expression
  \begin{align}
    \braket{\left(\Delta
        S^\alpha_0\left(t\right)\right)^2}&=\frac{1}{4}\frac{N}{N+1}.
  \end{align}
  The initial values enter in the equations of
  motion~\eqref{eq:04_EOM_classic} as well as in the autocorrelation function
  $\overline{S^z_0(t)S^z(0)}$ of the central spin.
\end{subequations}

Sampling the Gaussian fluctuations $\Delta\vec{\eta}(t)$ and the initial
values sufficiently well we determine the correlations by averaging over a
large number $M$ of runs at each instant $t$. The resulting data are shown in
Fig.~\ref{fig:2} labeled by ``Langevin 3''.

\subsection{Semiclassical results for zero field}
\label{ss:zero-field}

Here, we present results for the three approaches labeled ``Langevin 1'',
``Langevin 2'', and ``Langevin 3'' and compare them to the full quantum-mechanical DMRG results. The impact of the conservation of the total spin on
the semiclassical calculation is a particular focus.

\begin{figure}[tb]
  \centering
  \includegraphics[width=\columnwidth]{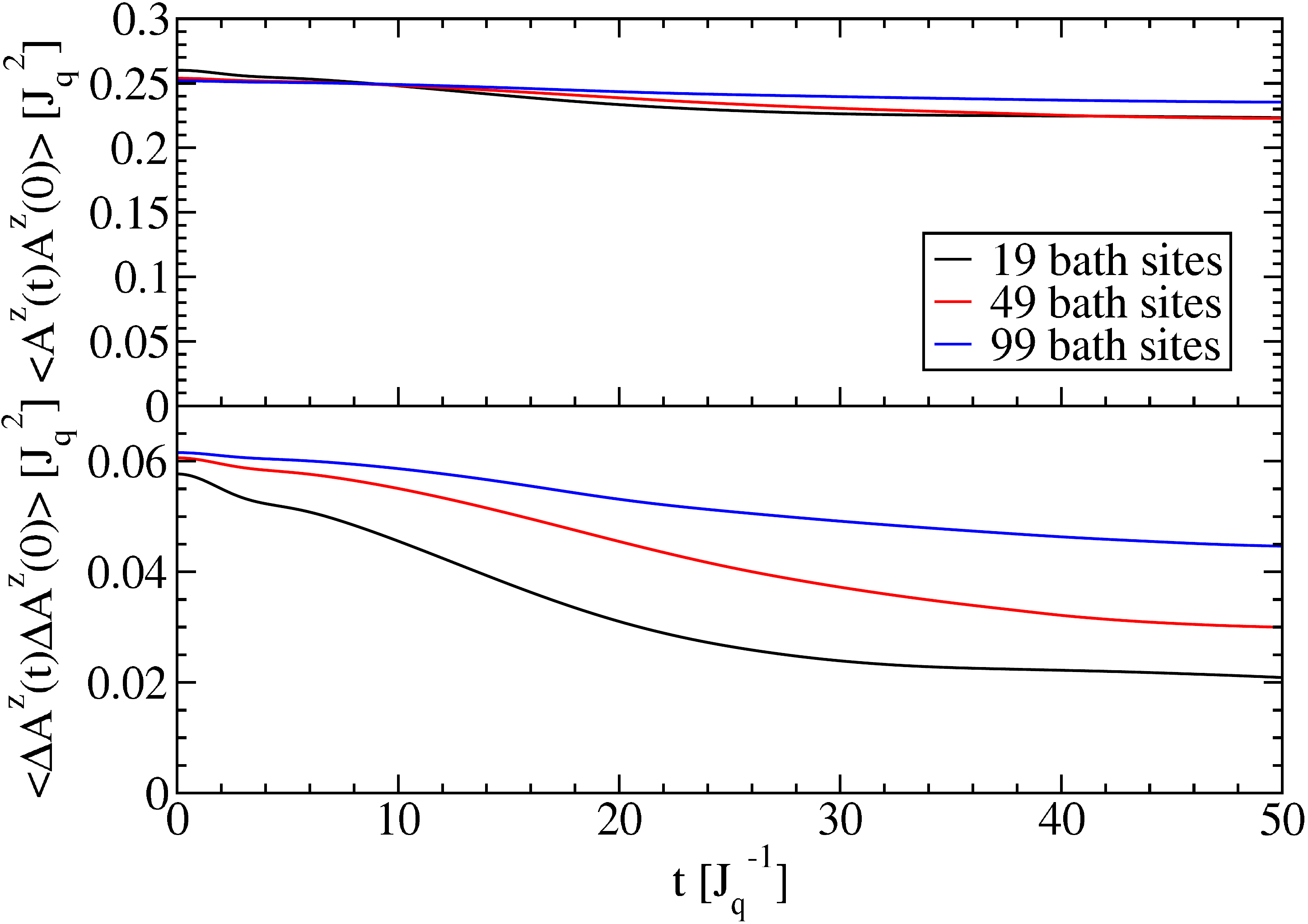}
  \caption{(Color online) (Top) Autocorrelation $\braket{A^z(t)A^z(0)}$
    as defined in Eq.~\eqref{eq:04_A_opt}. (Bottom) Separated fluctuating
    part $\braket{\Delta A^z(t)\Delta A^z(0)}$. Both are obtained by DMRG; the
    total discarded weight does not exceed $10\,\%$ at $t\approx 50/\Jq$, for
    further discussion see main text.}
  \label{fig:1}
\end{figure}

Before we consider the dynamics of the central spin we have to address the
autocorrelation of the Overhauser field $\vec{A}$ defined in
Eq.~\eqref{eq:04_A_opt}. This is mandatory because we need the correlations of
the noise as input for any semiclassical calculation. In the calculations
presented below, we will use the exact autocorrelation of the Overhauser field
for this purpose. Of course, one may object that no semiclassical calculation
is needed if we perform a DMRG calculation anyway. While this is true, we will
see that one understands the essentials of the central-spin dynamics better
from the semiclassical calculations.

In the upper panel of Fig.~\ref{fig:1}, the DMRG results for the total
autocorrelation of the Overhauser field including the constant part are
depicted. In the lower panel, the fluctuating part $\Delta\vec{A}$ is plotted.
The data are obtained from the DMRG implementation for the CSM involving a
purified initial state describing the disordered bath and the time evolution
based on the Trotter-Suzuki decomposition.\cite{fried06,Stanek2013} Beyond the
time range shown the total discarded weight starts exceeding 10\%.  The total
(accumulated) discarded weight comprises the sum of the discarded weight in
the reduced density matrix of all involved DMRG basis truncations up to the
given time including the DMRG buildup and the DMRG sweeps. The percentage of
total discarded weight implies that the deviations of generic expectation
values can be in the same order of about 10\%. But the increase is roughly
exponential as observed earlier, cf.\ Ref.~\onlinecite{Stanek2013}. This
means that the relative error at $t\approx 45/\Jq$ is $1\%$ and at $t\approx
40/\Jq$ it is $0.1\%$ and so on. Thus the results are very reliable except
close to the maximum times shown.

The lower panel demonstrates that the fluctuating part is indeed small
compared to the constant one. Moreover, it is decreasing so that for large
times one can expect that it vanishes completely and only the sizable
constant part remains. Compared to the results in Ref.~\onlinecite{Stanek2013} the definition of the Overhauser field is changed
because it includes the central spin itself now. This inclusion induces
slightly longer lasting correlations between the fluctuations that should
stabilize the autocorrelation of the central spin as well. As before in Ref.~\onlinecite{Stanek2013}, the autocorrelation converges towards
$\braket{A^z(t)A^z(0)}=\Jq^2/4$ for $N\rightarrow\infty$.

\begin{figure}[tb]
  \centering
  \includegraphics[width=\columnwidth]{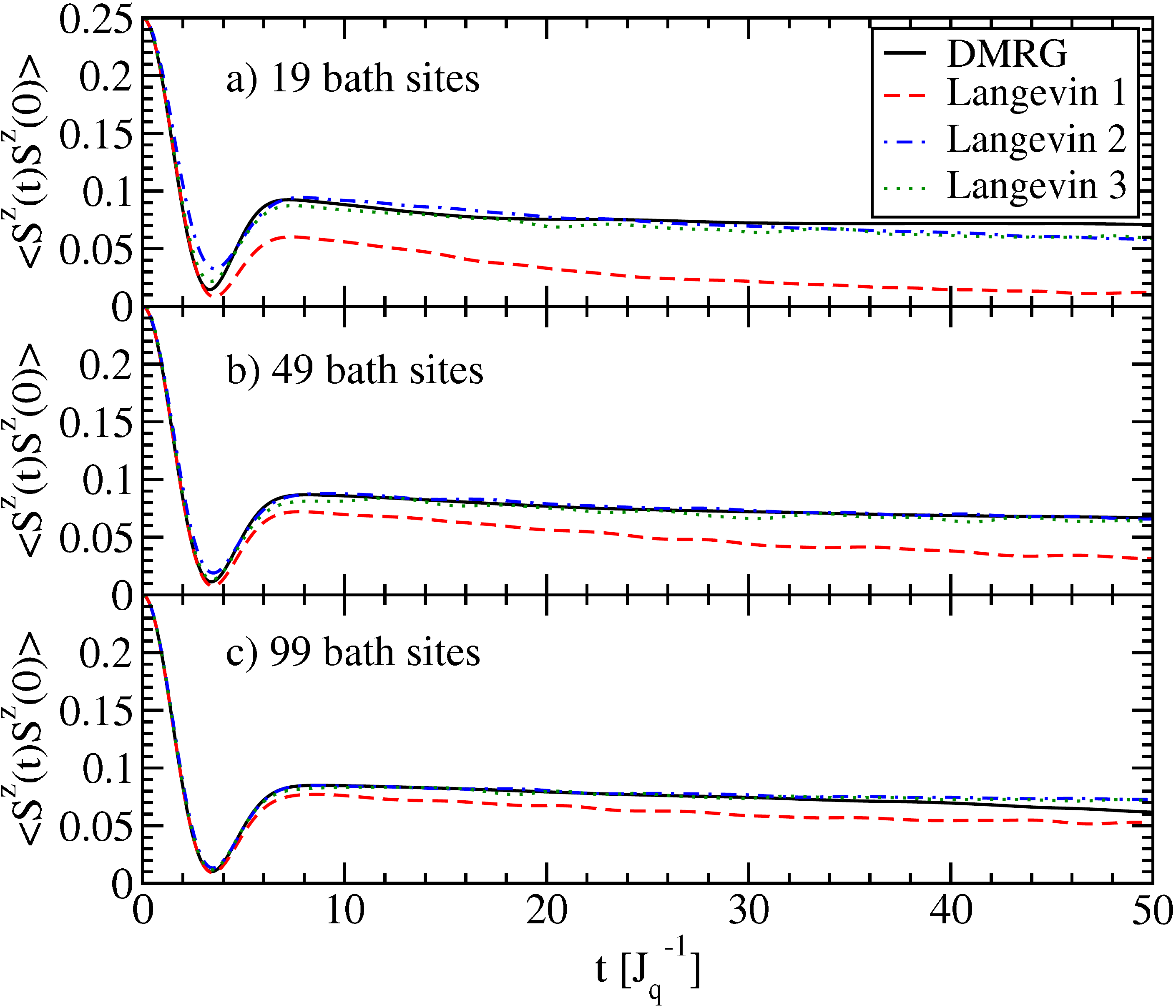}
  \caption{(Color online) Autocorrelation of the central spin up to intermediate time scales.
 In each panel, the results of the semiclassical calculations are compared to the corresponding DMRG result.
 All curves involving random noise have been obtained for $M=50\,000$ fluctuations, i.e., $M$ different
 time series $\vec{\eta}(t)$ and initial vectors are generated and the resulting central-spin evolutions
 are averaged.
 \label{fig:2}}
\end{figure}

For the simulation of the random noise, a large number $M$ of time series
$\vec{\eta}(t)$ and initial vectors are sampled. Technically, correlated noise
obeying a known autocorrelation $g(t)$ or $\Delta g(t)$ is generated from
Gaussian distributed random variables by means of the eigen decomposition of
the covariance matrix. The latter is obtained from the autocorrelations
$\braket{A^z(t)A^z(0)}$ or $\braket{\Delta A^z(t)\Delta A^z(0)}$ discretized
in time and taken from a DMRG calculation. The classical or quantum-mechanical
time integrations are carried out easily for each time series $\vec{\eta}(t)$
because we are only dealing with a two-level system or a classical
three-component vector. Finally, the average over the resulting $M$ time
evolutions is computed. The concomitant statistical relative errors are
estimated by $1/\sqrt{M}$.

The results for the autocorrelation $\braket{S^z_0(t)S^z_0(0)}$ of the central
spin are presented in Fig.~\ref{fig:2}. For each bath size, we compare the
DMRG result with the results obtained from the three semiclassical approaches
introduced in the previous sections. If the conservation of the total spin
is explicitly included, the central spin is treated on the operator level
(Langevin 2) and as a classical vector (Langevin 3). As expected, however, this
does not make any noticeable, statistically relevant difference for larger
baths.

In contrast, the difference between Langevin 1 (without conservation of spin)
and Langevin 2 or 3 (with conservation of spin) is significant. The data
clearly show the importance of treating conserved quantities properly. The
explicit conservation of the total spin leads to a substantial improvement of
the results, which decay slower than in Langevin 1 in agreement with DMRG. The
agreement with DMRG improves quickly with increasing system size. Up to the
times displayed, the agreement of Langevin 2 or 3 with the DMRG data is
excellent for the larger bath. We attribute the fact that for 99 bath sites
the DMRG data fall below the semiclassical result beyond about $t\approx
40\Jqt$ to the growing inaccuracy of the DMRG data for growing $t$.

In view of the above finding, the question arises whether the semiclassical
result stays close to the true quantum-mechanical result for all times. The
answer is ``no''. If the semiclassical computation is extended to long times
$t\gg 50\Jqt$ \emph{assuming} even that the Overhauser field autocorrelation
does not decrease further beyond $t=50\Jqt$ one finds a decrease of the
autocorrelation of the central spin down to a small value which is protected
by the conservation of the total spin. However, this fraction is small; it scales
like $1/N$ for $N\to\infty$, i.e., it is zero for infinite systems.

This is in contrast to the rigorously established behavior that a finite
spin-spin correlation persists for all times if the average coupling is
finite.\cite{uhrig14} This observation stems from Mazur's inequality. It turns
out that the conservation of the total spin is only one ingredient, but the
conservation of the total energy is another important prerequisite. Only the
energy conservation leads to a lower bound to the spin-spin correlation of the
central spin which does not vanish for $N\to\infty$ if the average coupling
remains finite. Thus we conclude that the semiclassical approach, even if it
is enhanced by the spin conservation, tends to fail for long times, at least
for zero external field.

\subsection{Semiclassical results for finite external field}

\begin{figure}[tb]
  \centering
  \includegraphics[width=\columnwidth]{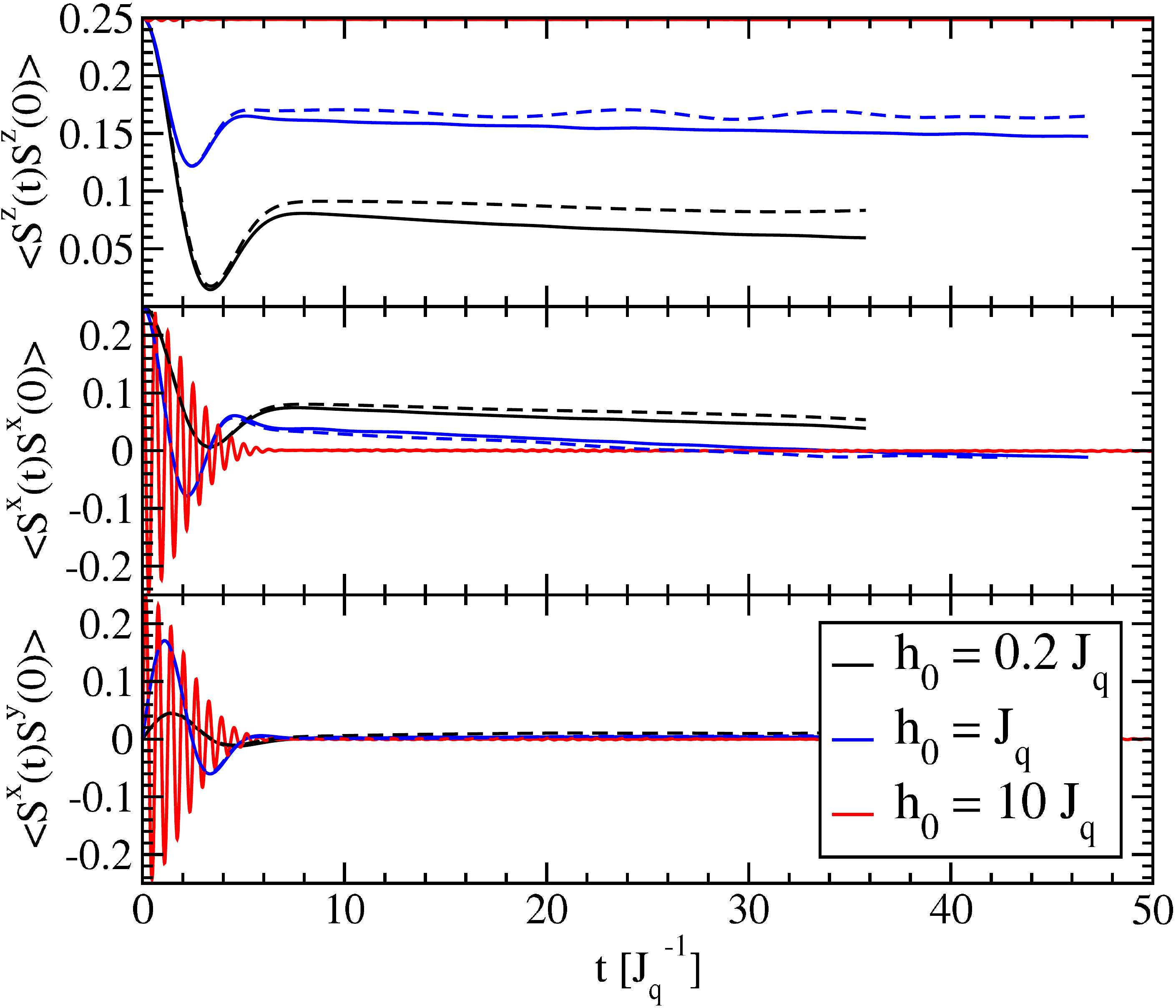}
  \caption{(Color online) Nonvanishing autocorrelations
    $\overline{S_0^\alpha(t)S^\beta_0(0)}$ of the central spin for $N=99$ bath
    spins and various strengths of the external magnetic field. The solid
    lines represent the solution of the semiclassical model ``Langevin 1''
    while the dashed lines are the quantum results calculated with DMRG. The
    results are shown up to the time where the total discarded weight of the
    DMRG calculations exceeds $10\,\%$, cf.\ discussion in Sect.~\ref{ss:zero-field}. The semiclasscial simulation involved $M=100\,000$
    random time series. }
  \label{fig:3}
\end{figure}

The semiclassical model~\eqref{eq:04_HSC} can also be used for finite external
fields. For brevity, we restrict ourselves to the simple semiclassical
model~\eqref{eq:04_HSC} (Langevin 1) without conservation of the total
spin. The cases of a weak field $h_0=0.2\,\Jq$, an intermediate field
$h_0=\Jq$, and a strong field $h_0=10\,\Jq$ applied to the central spin are
investigated. In these cases, the mean value of the Gaussian fluctuations in
$z$ direction acquires the role of the finite external field
\begin{align}
  \overline{\eta_z\left(t\right)}&=h_0,
\end{align}
while all other mean values remain zero. The nonvanishing correlation
functions $\braket{A^z(t)A^z(0)}$, $\braket{A^x(t)A^x(0)}$, and
$\braket{A^x(t)A^y(0)}$ of the Overhauser field are calculated with DMRG. The
results are determined up to the time where the total discarded weight exceeds
$10\,\%$. More details on the implied accuracy can be found in Sec.~\ref{ss:zero-field}. They serve as input for the cylindric correlation
functions $g_{zz}(t)$, $g_{xx}(t)$, and $g_{xy}(t)$, respectively.  The
nonzero autocorrelations of the central spin are plotted in Fig.~\ref{fig:3}
for a bath of $N=99$ spins.

A very nice agreement between the semiclassical (solid lines in Fig.~\ref{fig:3}) and the quantum data (dashed lines) is found in the strong-field
regime where the dynamics is dominated by the Larmor precession of the central
spin. For low and intermediate fields, a certain mismatch between both
descriptions is always present. In general, the quantum and the semiclassical
approach agree remarkably well. Note that the discrepancies are more
pronounced in the spin direction parallel to the external field than in the
perpendicular direction.

We attribute the observed mismatch to the missing conservation of the total
$z$ component of the spin. We refrain here from discussing the possible
improvement by its inclusion because such calculations would still need the
input of the bath correlations. Instead, we will show below that a completely
classical simulation is more efficient and more reliable.

\subsection{Limitations of the semiclassical approach}

We achieved a significant improvement in the semiclassical description of the
CSM by incorporating the conservation of the total spin explicitly. For finite
external fields, the incorporation of the $z$ component of the total spin is
not strictly required because a good agreement in spin directions
perpendicular to the external field is already achieved without this
conservation. However, two essential disadvantages of the semiclassical
approach remain.

First, it relies on an additional method providing the correlations of the
random noise of the Overhauser field. Hence, accessible time scales are
limited by this additional method and almost no resources are saved.

Second, the semiclassical approach does not respect the energy conservation of
the CSM
\begin{align}
  0&\stackrel{!}{=}\frac{\mathrm{d}}{\dt}H_\mathrm{sc}=
  \frac{\mathrm{d}\vec{\eta}\left(t\right)}{\dt}\cdot\vec{S}_0\neq 0.
  \label{eq:04_dHdt}
\end{align}
The energy is conserved in the quantum model where the state of the bath
depends on the state of the central spin such that $H$ is constant. The
relevant important backactions effects are \emph{not} included in the
semiclassical picture so that the energy conservation is lost. This may be
repaired by some clever incorporation of the energy conservation in a similar
way to what we did for the spin conservation. However, the task appears to be
complicated and in view of the first, remaining caveat less attractive.

These considerations lead us to the next section, where a completely classical
simulation is compared to the quantum-mechanical results.

\section{Classical Equations of Motion}
\label{sec:classical}

In the previous section, we illustrated the importance of the proper treatment
of conserved quantities. This conclusion is supported further by the rigorous
bounds for the correlations of the central spin found very
recently.\cite{uhrig14} Thus, searching for a computationally simple approach
to the CSM it is natural to think of classical calculations
\cite{Merkul2002,Erling2004,AlHass2006,Chen2007,Coish2007} because the
classical model has the same conserved quantities as the quantum-mechanical
one.

In addition, the norm of the commutators of the Overhauser field vanish
relative to the norm of the Overhauser field itself for an infinite bath
$N\to\infty$.\cite{Stanek2013} Note that in quantum dots $N$ is of the order
of $10^5$. Thus it is well justified to treat the Overhauser field
classically. We showed in the previous section that the dynamics of the
central spin can also be determined by considering it as classical vector due
to the linearity of the corresponding equations of motion. Only for the
backaction of the central spin on the individual bath spins it is not evident
if a classical treatment resembles the quantum-mechanical one.

A qualitative argument in favor of the agreement of the classical and the
quantum-mechanical approach is the separation of time scales: the central spin
precesses faster by a factor of about $\sqrt{N}$ than a single bath spin. So
the single bath spin is exposed to a long-time average of the central-spin
dynamics. Such an average is believed to behave classically.

It is the purpose of the present section to study quantitatively to what
extent the quantum-mechanical dynamics and the classical one agree. The
relevant classical $3(N+1)$ equations of motion (EOMs) read
\begin{subequations}
  \label{eq:05_EOM}
  \begin{align}
    \frac{\mathrm{d}}{\dt}\vec{S}_0 &=\vec{A}\times\vec{S}_0-\vec{h}_0\times\vec{S}_0 \\
    \frac{\mathrm{d}}{\dt}\vec{S}_i &=J_i\vec{S}_0\times\vec{S}_i,
  \end{align}
\end{subequations}
where $i\in\left\{1,2,\ldots,N\right\}$ and $\vec{h}_0$ is the external field
applied to the central spin. The classical Overhauser field is defined by
\begin{align}
  \vec{A}&:=\sum\limits_{i=1}^N J_i\vec{S}_i.
\end{align}
The Overhauser field is a vector in $\mathds{R}^3$ and not an operator anymore
as in the quantum model. The equations \eqref{eq:05_EOM} are derived in Ref.~\onlinecite{AlHass2006} as a time-dependent mean-field approximation to the
quantum model.

It is easily verified that the total energy
\begin{subequations}
  \begin{align}
    E&=\vec{A}\cdot\vec{S}_0 \\
    \Rightarrow \frac{\mathrm{d}E}{\dt}&=\left(\frac{\mathrm{d}\vec{A}}{\dt}
    \right) \cdot\vec{S}_0+\vec{A}\cdot\left(\frac{\mathrm{d}\vec{S}_0}{\dt}
    \right) =0
  \end{align}
\end{subequations}
is conserved because of the properties of the outer product.

The set~\eqref{eq:05_EOM} of coupled EOMs is solved best using standard
numerical routines such as Runge-Kutta integration. Here, we stick to the
adaptive Runge-Kutta-Fehlberg method which is part of the \textsc{GNU
  Scientific Library} (GSL).\cite{GSL2009} As in Sect.~\ref{ss:centrspin-classvec}, the initial values for all spins $\vec{S}_i(t)$
at $t=0$ are chosen from a Gaussian distribution with vanishing mean
value. The variance is given from the quantum-mechanical expectation values
for disordered $S=1/2$:
\begin{align}
  \overline{S^\alpha_i(0)S^\alpha_i(0)}&=\frac{1}{4}.
\end{align}
By the numerical integration, one obtains the time evolution of all spins
$\vec{S}_i$. The desired autocorrelations of the central spin are calculated
by averaging over a large number $M$ of random initial configurations. This
corresponds to the investigation of a completely unpolarized system at
infinite temperature. It would be straightforward to implement analogous
calculations for (partially) polarized baths.

In our numerical implementation, we checked the conservation of the energy and
of the total spin explicitly to verify the correctness of the code. The energy
in the studied time interval $t\in[0,1000\,\Jqt]$ is conserved within
$10^{-6}$, which also corresponds to the step size of the Runge-Kutta
integration. On the same time scale, the total momentum is conserved within
$10^{-12}$. Decreasing the step size of the Runge-Kutta method did not lead to
a significant improvement so that we used $10^{-6}$ as standard value
throughout.

\subsection{Classical results for zero field}

\begin{figure}[tb]
  \centering
  \includegraphics[width=\columnwidth]{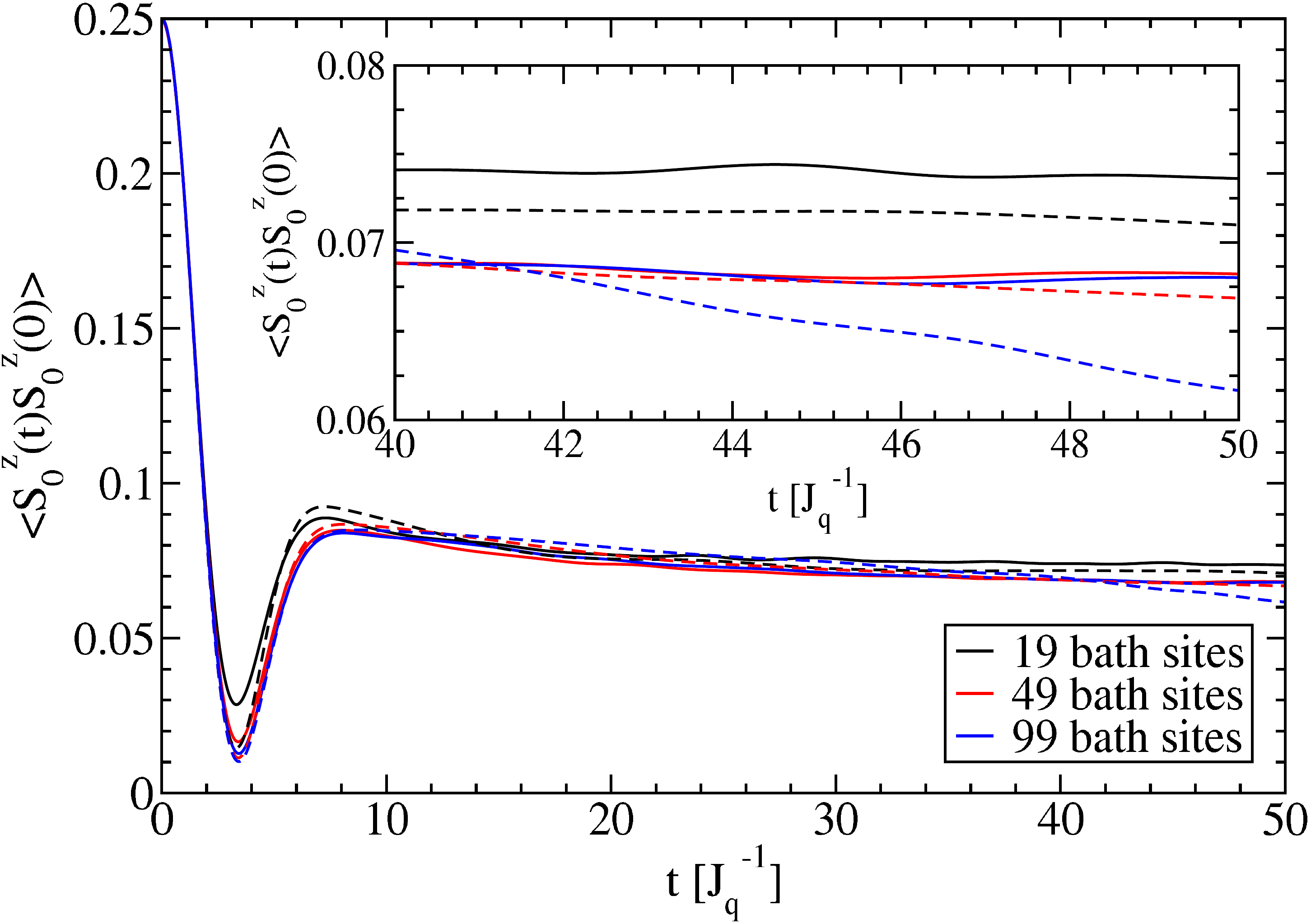}
  \includegraphics[width=\columnwidth]{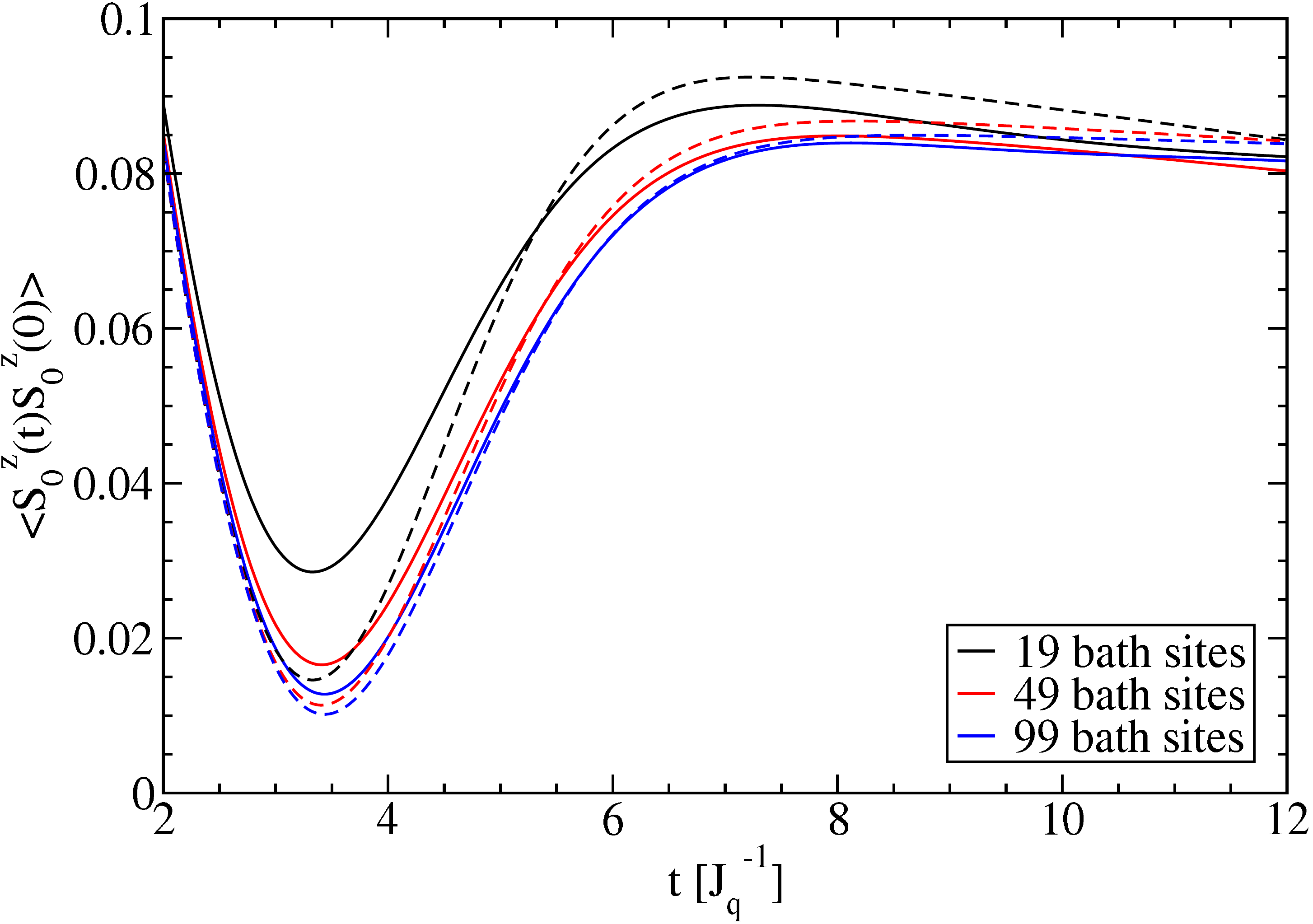}
  \caption{(Color online) (Top) Autocorrelation function of the central
    spin for zero field up to intermediate time scales. In addition to the
    classical solution (solid lines), the DMRG results of the corresponding
    quantum model are plotted (dashed lines). The inset displays a
    magnification for $t\ge 40\,\Jqt$. On these time scales, the DMRG result
    for $N=99$ appears to be slightly inaccurate, which we attribute to the
    total discarded weight. The results for the classical EOMs are averaged
    over $M=1\,000\,000$ random initial configurations of the spin vectors. The
    lower panel displays a magnification of the short-time behavior
    illustrating that the classical data and the quantum-mechanical DMRG data
    agree better and better for larger and larger spin baths.}
  \label{fig:4}
\end{figure}

At first, we study the case without external field. The results for the
autocorrelation of the central spin are plotted in Fig.~\ref{fig:4}. Overall,
the classical solutions (solid lines) and the DMRG results (dashed lines)
agree nicely up to intermediate times $t=50\,\Jqt$. The minimum close to
$t\approx4\,\Jqt$ is not correctly captured by the classical solution if the
bath size is small, cf.\ lower panel. However, fast convergence with $N$ is
observed so that only a marginal difference between the classical and the
quantum results remains for a moderate number of $N=99$ bath spins. Of course,
an absolute agreement cannot be expected since both quantum and classical
description are distinct. Hence the observed agreement between the two
approaches is already remarkable.

After the plateau of the autocorrelation has emerged beyond $t\approx 7\Jqt$,
a good agreement between classical and quantum results persists, in particular
for larger spin baths. It supports the idea that the large number of
interaction partners and the clear separation of energy scales makes the
quantum dynamics very close to the classical one.

The drop in the DMRG result for $N=99$ bath spins beyond $t\ge40\,\Jqt$ is to
be attributed to numerical inaccuracies because the total discarded weight is
close to 10\% on the respective time scale. More details on accuracy can be
found in Sect.~\ref{ss:zero-field}.

\begin{figure}[tb]
  \centering
  \includegraphics[width=\columnwidth]{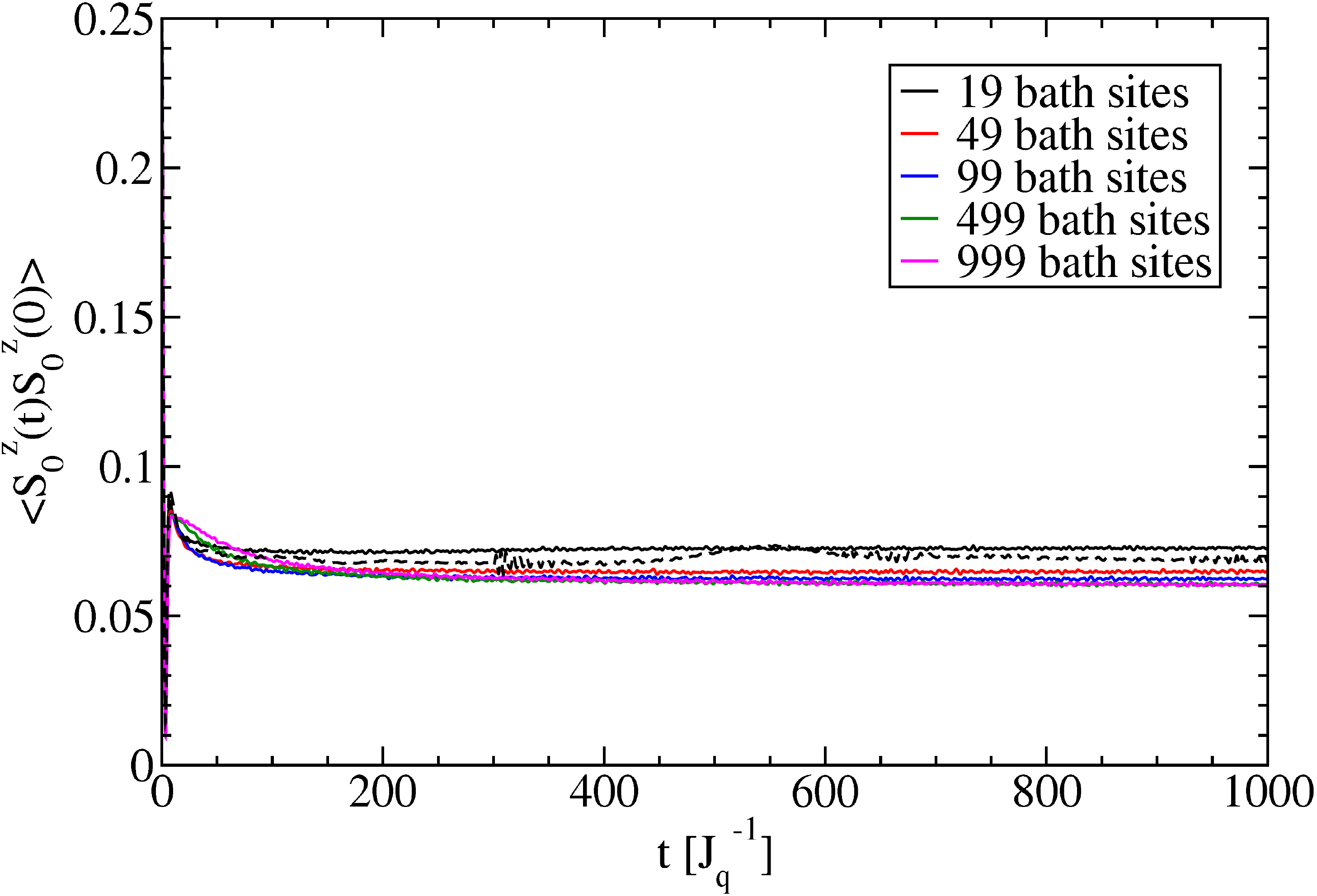}
  \includegraphics[width=\columnwidth]{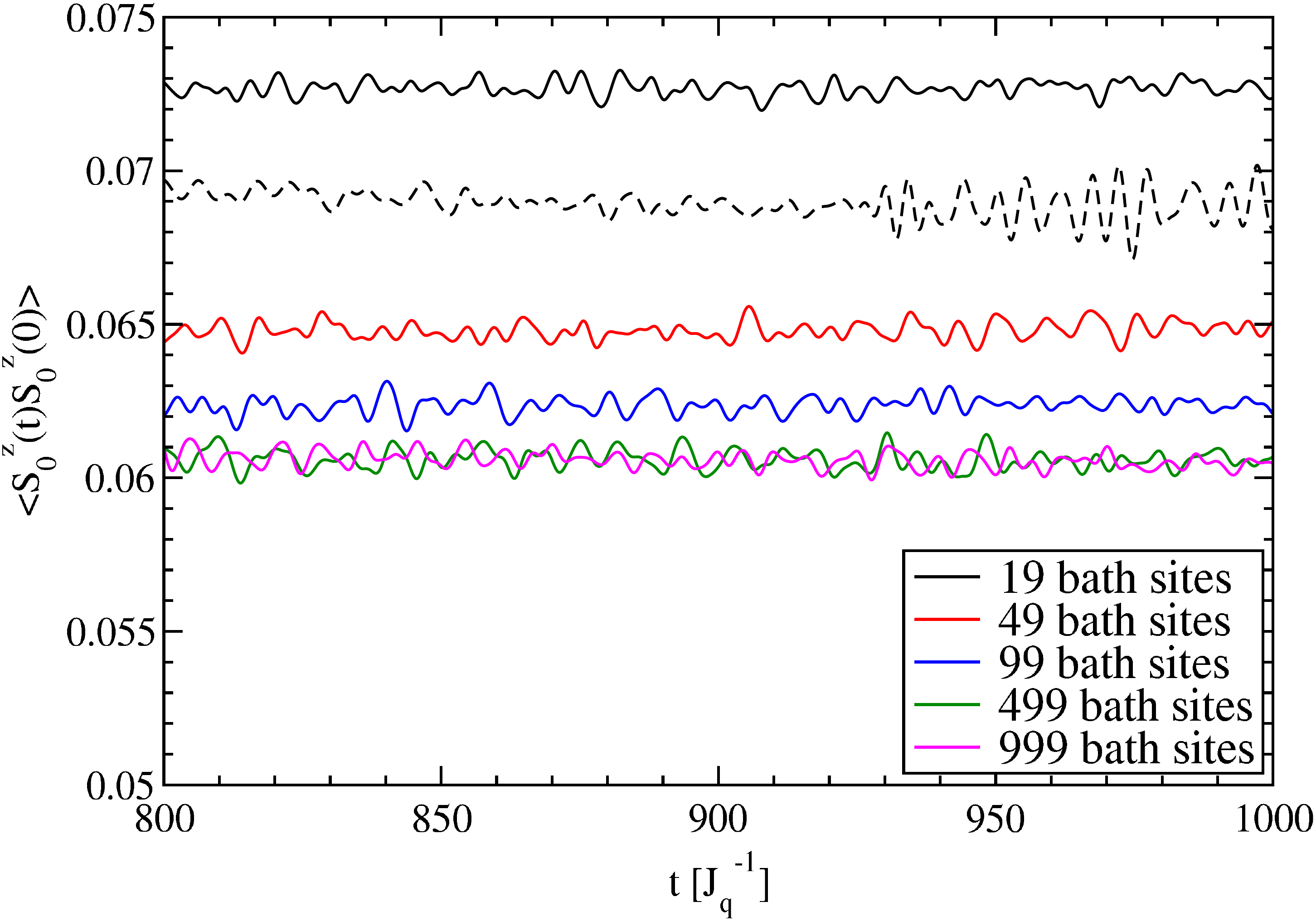}
  \caption{(Color online) (Top) Long-time behavior of the
    autocorrelation of the central spin obtained from the classical EOMs for
    up to $N=999$ bath spins. The lower panel displays a magnification for
    larger times $t\ge 800\,\Jqt$ where the curves fluctuate around a constant
    finite value. The black dashed curve is obtained from the Chebyshev
    expansion for $N=19$ bath spins.\cite{Hackma2013a,Hackma2014} The results
    for the classical EOMs are averaged over $M=1\,000\,000$ random initial
    configurations of the spin vectors.}
  \label{fig:5}
\end{figure}

Next, we address the behavior at long times. The key problem is that we do not
have reliable quantum data available for long times. A first impression is
provided in Fig.~\ref{fig:5} for a bath of $N=19$ spins for which times up to
$t=1000\Jqt$ are accessible by the Chebyshev expansion.\cite{Hackma2014} Only
bath sizes $\approx 20$ can be tackled by the Chebyshev expansion. For large
values of $t$, all curves acquire a plateau value which depends on the actual
bath size. The persisting plateau value is generally identified as the
nondecaying fraction of the autocorrelation.\cite{Fariba2013,Fariba2013a}
From the lower panel of Fig.~\ref{fig:5}, we see that the value of the
nondecaying fraction converges well with diverging system size $N$. We recall
that in real quantum dots the number of bath spins exceeds $10^5$ easily. By
mathematically rigorous bounds the existence of nondecaying fractions has
been established only recently for the quantum CSM, provided the mean coupling
$\overline{J}:=\lim_{N\to\infty} N^{-1}\sum_{i=1}^NJ_i$ does not
vanish.\cite{uhrig14} Estimates of the nondecaying fractions for the
classical CSM can be found in the literature.\cite{Merkul2002,Chen2007}

\begin{figure}[tb]
  \centering
  \includegraphics[width=\columnwidth]{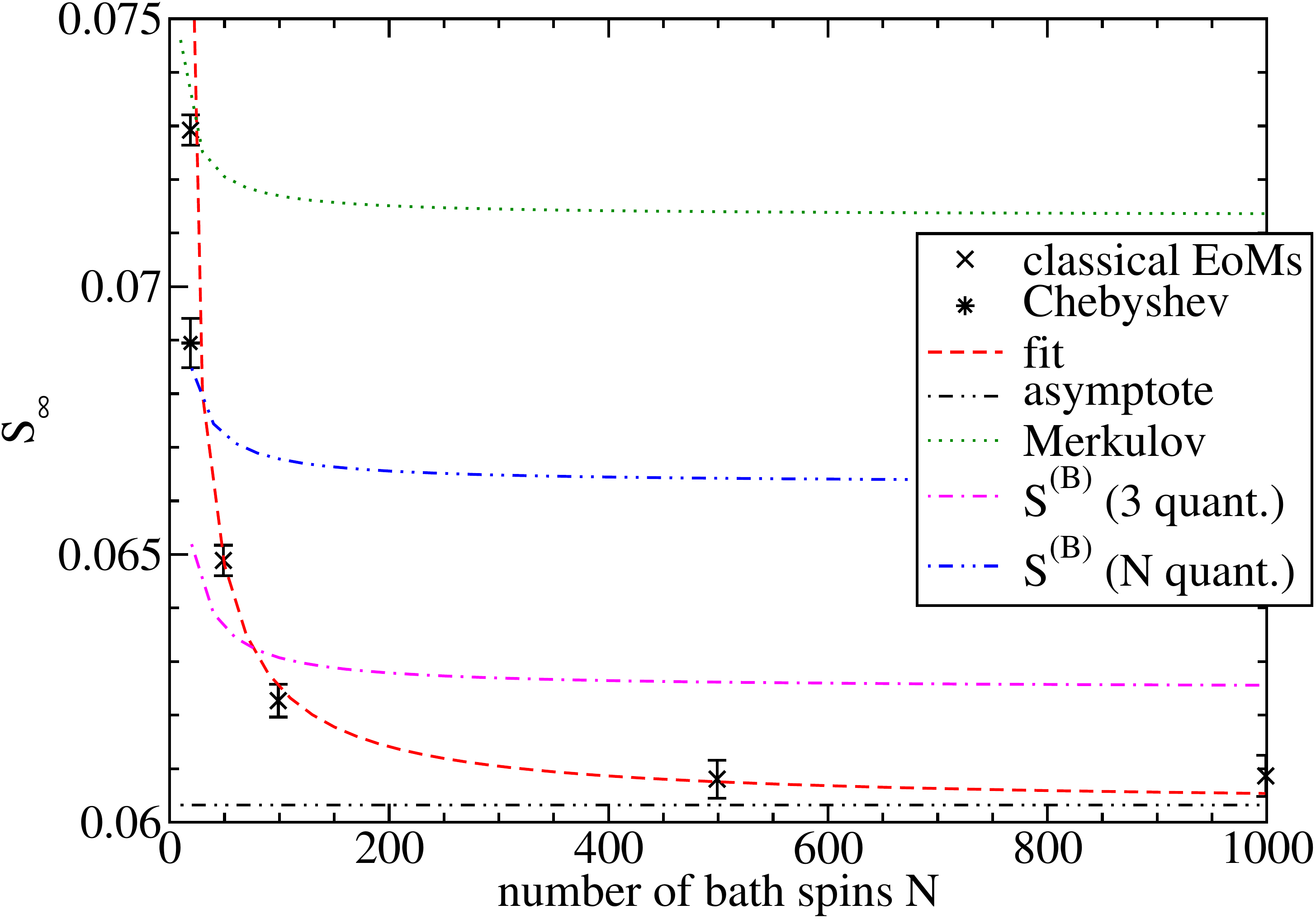}
  \caption{(Color online) Nondecaying fraction $S_\infty$ of the
    autocorrelation of the central spin (symbols) vs the number of bath
    spins $N$. The long-time results of the classical EOMs fluctuate around a
    constant value (see lower panel of Fig.~\ref{fig:5})), the mean values for
    $t=500$--$1000\,\Jqt$ are plotted. The error bars depict the standard
    deviation. The red curve represents the fit function
    $f(N)=3a(N/b+1)/(N/b-1)$ with $a=0.0201$ and $b=1.79$. To show the
    asymptotics, we plot the $N=\infty$ limit
    $f(N)\xrightarrow{N\rightarrow\infty}3a$ as determined by the fit (blue
    curve). The dotted (green) curve depicts the estimate from
    Ref.~\onlinecite{Merkul2002} for our set of coupling constants. In
    addition, the quantum-mechanical estimates for $S_\infty$ in the quantum
    model are depicted.\cite{uhrig14}}
  \label{fig:6}
\end{figure}

We want to elucidate the behavior of the nondecaying fraction resulting from
our classical simulation and compare it with the quantum-mechanical estimates
as they are provided by the formulas in Ref.~\onlinecite{uhrig14}. This is
done in Fig.~\ref{fig:6} where we plot the numerically determined values
$S_\infty$ for the nondecaying fraction, that is
\begin{align}
  S_\infty&:=\overline{S^z_0(t\rightarrow\infty)S^z_0(0)}.
  \label{eq:05_x}
\end{align}
This quantity has been estimated by Merkulov \textit{et al.} in Ref.~\onlinecite{Merkul2002} by a formula that depends only on the following
combination of the mean coupling $\overline{J}$ and the mean square coupling
$\overline{J^2}$
\begin{equation}
  \frac{\overline{J}^2}{\overline{J^2}-\overline{J}^2} =3\frac{N+1}{N-1}
  \label{eq:merkul}
\end{equation}
where the right-hand side applies to the couplings \eqref{eq:02_Ji} we are
studying here. The formula of Merkulov \textit{et al.} results are shown as the dotted
(green) line in Fig.~\ref{fig:6}. It is clearly deviating from the numerical
result by about 20\%. By allowing the constants in \eqref{eq:merkul} to vary, we introduce the fit function
\begin{equation}
  f(N)=\frac{3a(N/b+1)}{N/b-1}
\end{equation}
which describes the numerical data well, see the dashed (red) curve in Fig.~\ref{fig:6}, but it is of empirical value only.

Since we are ultimately interested in the comparison to the quantum case we
include the quantum-mechanical estimates $S^{\textrm{(B)}}$ from Ref.~\onlinecite{uhrig14} which consider three or $N$ conserved quantities. Note the
excellent agreement of the estimate from $N$ conserved quantities with the
Chebyshev result for $N=19$ bath spins (star symbol). Thus, we assume that the
estimate remains as close as this to the true nondecaying fraction. Then, we
have to deduce that the classical simulation \emph{underestimates} the quantum-mechanical result by about 10\%. This is interesting because it is \textit{a priori}
unclear whether the classical correlation decays more or less than the quantum
correlation. One would have thought that quantum fluctuations reduce a
persisting correlation below the classical value. However, it seems that the
classical phase space imposes less restrictions on the central-spin dynamics
than the quantum-mechanical Hilbert space. Certainly, the quantitative
comparison of the classical and the quantum-mechanical nondecaying fraction
deserves further investigation.

Summarizing, the quantum-mechanical dynamics of the electron spin in zero
magnetic field is almost quantitatively described by the classical simulation
up to intermediate time scales. As a general trend, the agreement between
quantum and classical dynamics improves upon increasing bath size.

On very long-time scale, the issue is not completely clear. Classical and
quantum-mechanical results display a significant nondecaying fraction. There
is evidence that the nondecaying fractions are larger in the quantum-mechanical CSM than in the classical CSM. So the entanglement between central
spin and bath appears to protect the coherence, at least partially.

In essence, our observations agree with results by Coish~\textit{et
  al.}.\cite{Coish2007} They compared the quantum solution with the
corresponding classical solution for a single initial state and found that the
dynamics is essentially classical up to a certain time. Beyond that time,
quantum fluctuations have to be taken into account. However, they did not
study the average over all initial conditions as we do. Furthermore, all
couplings in their study were homogeneous $J_i=J$ and the expectation values
of the observables were calculated on the mean-field level. A numerical
solution of the full set of EOMs was not considered.

The classical equations of motion \eqref{eq:04_EOM_classic} can also be viewed
as the equations for a time-dependent mean-field
approximation.\cite{AlHass2006} As such this approximation turns out to be
fairly crude.\cite{Dobrov2003,AlHass2006} Thus, we conclude that the average
over the manifold of classical trajectories must be the key improvement in our
classical approach. The assumption to use Gaussian distributed random initial
spin orientations which are compatible with the initial quantum-mechanical
expectation values appears to be the appropriate choice.

Al-Hassanieh \textit{et al.} also start from random initial spin orientations,
which are compatible with the initial density matrix.\cite{AlHass2006} In this
respect, both approaches are similar. Then, however, they derive a set of
differential equations for the approximate temporal evolution of the density
matrix. Finally, this set is integrated to describe the dynamics of the
central spin. The authors emphasize, that their approach does not amount up to
the integration of semiclassical equations of motion. In contrast, we employ
the classical equations of motion \eqref{eq:04_EOM_classic}, but with suitably
weighted averages over various initial conditions. The numerical effort
required in both approaches appears to be very similar because finally, a set
of ordinary differential equations has to be integrated over time. Their
number is of the order of the number of spins considered.

\subsection{Classical results for finite external field}

\begin{figure}[tb]
  \centering
  \includegraphics[width=\columnwidth]{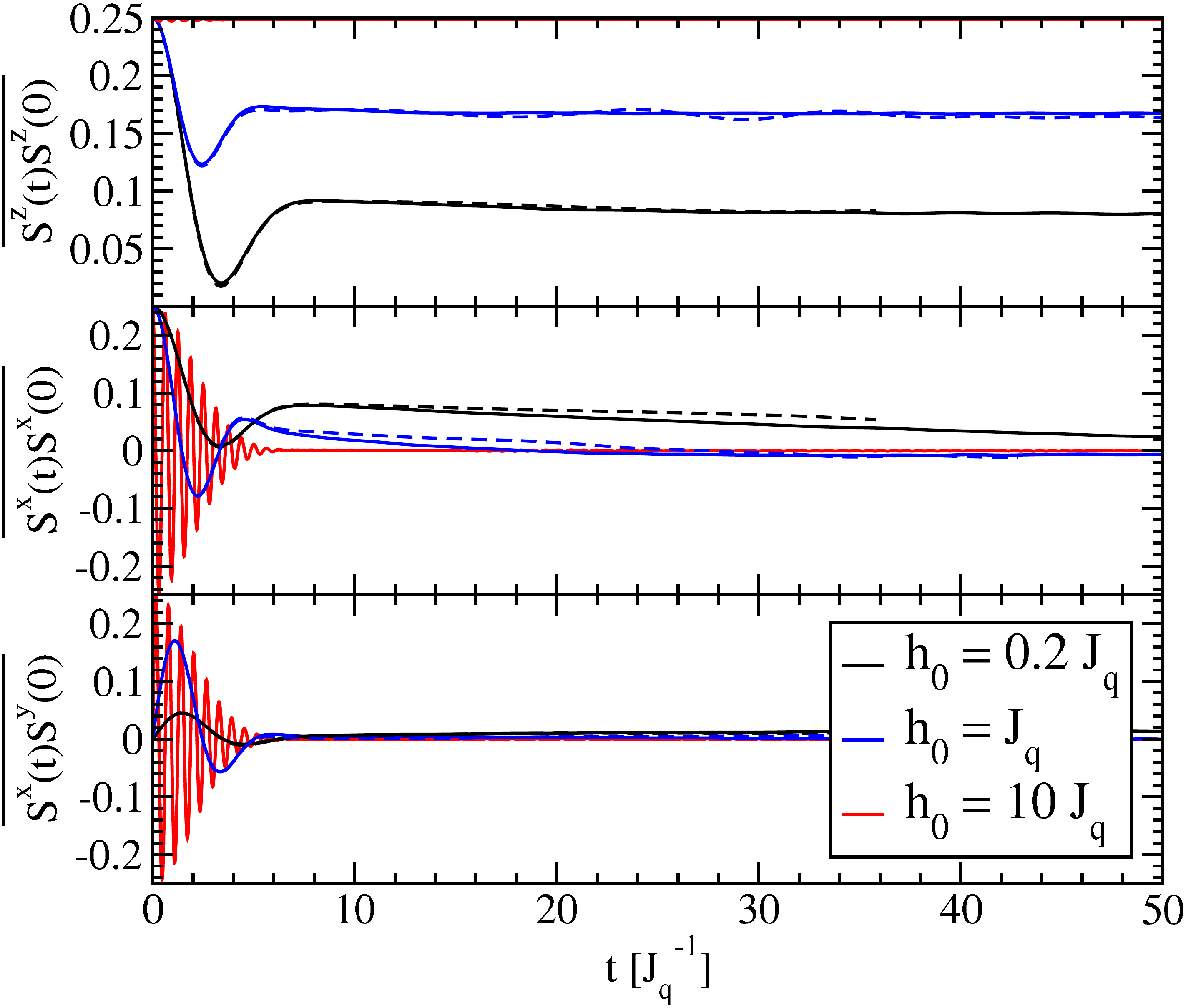}
  \caption{(Color online) Nonvanishing autocorrelations
    $\overline{S_0^\alpha(t)S^\beta_0(0)}$ of the central spin for $N=99$ bath
    spins and various strengths of the external magnetic field. The solid
    lines represent the classical simulation while the dashed lines are the
    quantum results calculated by DMRG. The latter are shown up to the time
    where the total discarded weight exceeds $10\,\%$. (The implied accuracy
    is discussed in detail in Sect.~\ref{ss:zero-field}.)  Where they are
    not visible they lie underneath the classical curves. All classical curves
    are obtained by averaging over $M=1\,000\,000$ random initial configurations
    of the spin vectors.}
  \label{fig:7}
\end{figure}

In this section, we consider the classical simulations in presence of a finite
external field. For simplicity, it is applied along the $z$ direction
$\vec{h}_0=(0\ 0 \ h_0)^\top$. According to the EOMs in Eq.~\eqref{eq:05_EOM},
the external field is solely applied to the central spin due to the three
orders of magnitude between the electronic and the nuclear gyromagnetic ratio.
As before, the case of a weak field $h_0=0.2\,\Jq$, an intermediate field
$h_0=\Jq$, and a strong field $h_0=10\,\Jq$ is investigated. The following
results comprise the behavior up to intermediate times. In Fig.~\ref{fig:7},
the nonvanishing autocorrelations of the central spin are plotted for the
three regimes of the external field for a fixed bath size of $N=99$ spins. The
solutions of the classical EOMs (solid lines) are compared to the behavior
found in the quantum model (dashed lines of the same color/shading).

The essential dynamics of the central spin is captured by the classical
EOMs. For weak and intermediate values of $h_0$, a noticeable influence of
quantum fluctuations persists which leads to quantitative deviations between
the classical and the DMRG results on intermediate time scales. Still, for
large baths one observes qualitatively the same behavior. In the spin
direction parallel to the external field, hardly any influence of the quantum
fluctuations is discernible. Perpendicular to $h_0$, the classical
autocorrelations decay slightly faster than their quantum counterparts.

We think that these findings imply that the very general argument based on
spin path integrals that the classical behavior describes the quantum-mechanical one for large baths \cite{Chen2007} must be regarded with some
caution. On the basis of our findings it is obvious that the agreement of both
behaviors depends on the parameter regime, in particular on the size of the
applied magnetic field. Thus we presume that the convergence of classical and
quantum-mechanical regime is not uniform: up to a given time one can find a
system size $N$ above which the classical and the quantum curves agree
well. However, for small magnetic fields and for fixed system size $N$, there is a
time beyond which both approaches behave differently.

\begin{figure}[tb]
  \centering
  \includegraphics[width=\columnwidth]{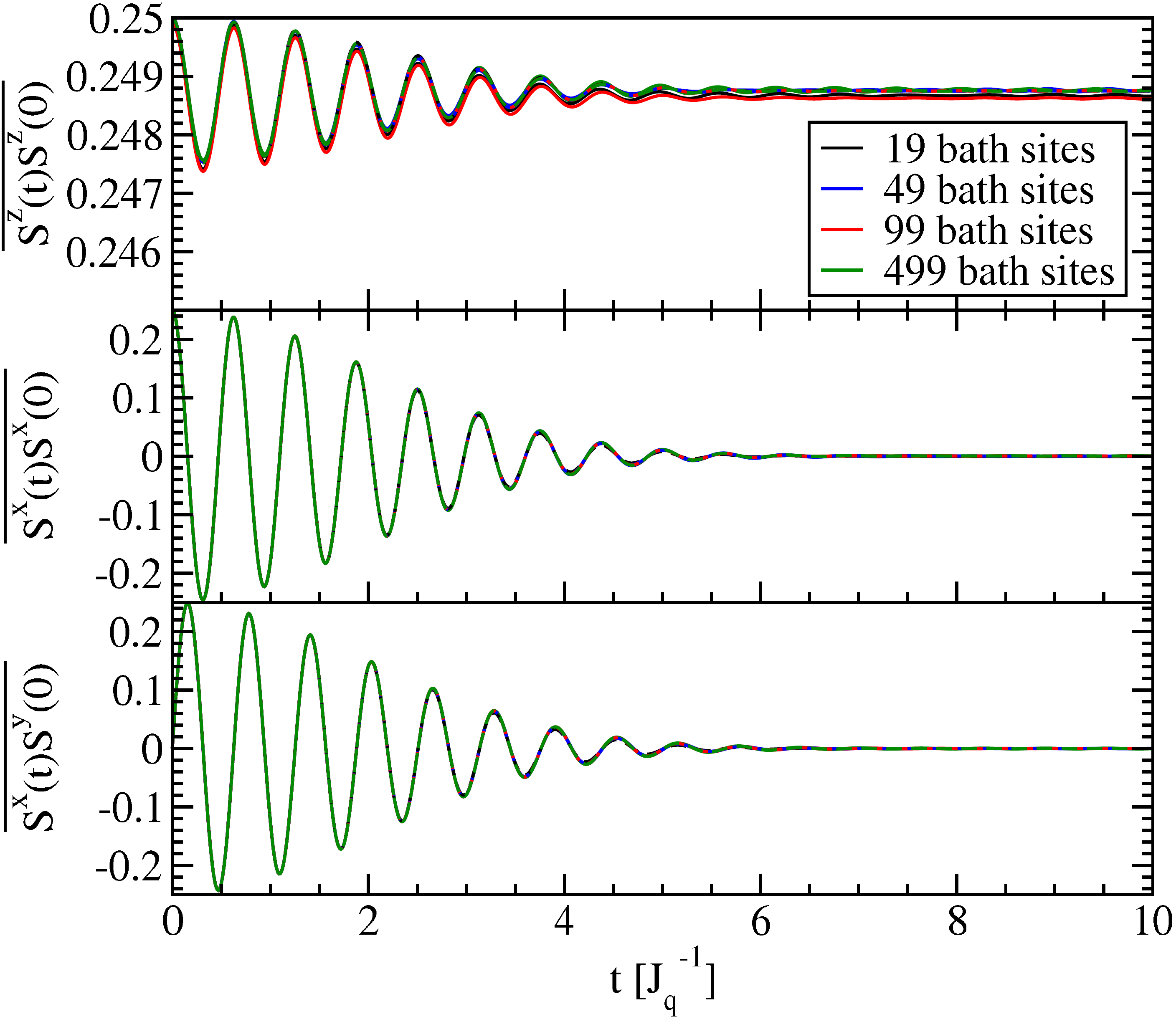}
  \caption{(Color online) Nonvanishing autocorrelation functions
    $\overline{S_0^\alpha(t)S^\beta_0(0)}$ of the central spin for a strong
    external field $h_0=10\, J_\mathrm{q}$. No longer times are depicted
    because no visible dynamics takes place beyond $t\approx 10 \Jqt$. The
    solid lines represent the solution of the classical EOMs while the dashed
    lines are the quantum results calculated with DMRG. All classical curves
    are obtained by averaging over $M=1\,000\,000$ random initial configurations
    of the spin vectors.}
  \label{fig:8}
\end{figure}

Remarkably, for strong external field $h_0\gg \Jq$, the quantum and classical
solutions cannot be distinguished. Generally, the classical EOMs indeed
capture the crossover from the weak to the strong field regime visible in the
DMRG data in Fig.~\ref{fig:7}. In the strong field regime, the dynamics in the
central-spin model is classical as demonstrated in Fig.~\ref{fig:8}. Note as
well that in this regime, the dependence on the number $N$ of bath spins is
extremely weak, i.e., the convergence with system size is rapidly
achieved. The dynamics is dominated by the Larmor precession induced by the
strong external field.

We point out that DMRG also exhibits an extremely good performance in the
strong field regime: the external field suppresses the relaxation of the
central spin, i.e., its entanglement with the bath degrees of freedom. Hence,
the number of important states to be kept by the DMRG increases significantly
more slowly with $t$ than for low or zero magnetic field. This implies an
important reduction of the total discarded weight at fixed number of kept
states. In consequence, the code runs faster and/or much larger times can be
reached. This can be made quantitative by the small total discarded weight
which does not exceed $\mathcal{O}(10^{-4}$--$10^{-3})$ even on long time
scales $t\approx 100\Jqt$. On the methodological level, this is a key
observation for the use of DMRG in the description of qubits.

At this stage, we are in the position to address one of the essential question
of spin decoherence. We consider the standard relaxation times. The so-called
spin-lattice relaxation time $T_1$ quantifies the longitudinal relaxation
parallel to the external field. From Fig.~\ref{fig:8}, it is obvious that no
substantial longitudinal relaxation takes place, see the scale of the
uppermost panel depicting the $S^zS^z$ correlation. Thus $T_1=\infty$ holds
for the CSM in a sufficiently strong magnetic field.

\begin{figure}
  \centering
  \includegraphics[width=\columnwidth]{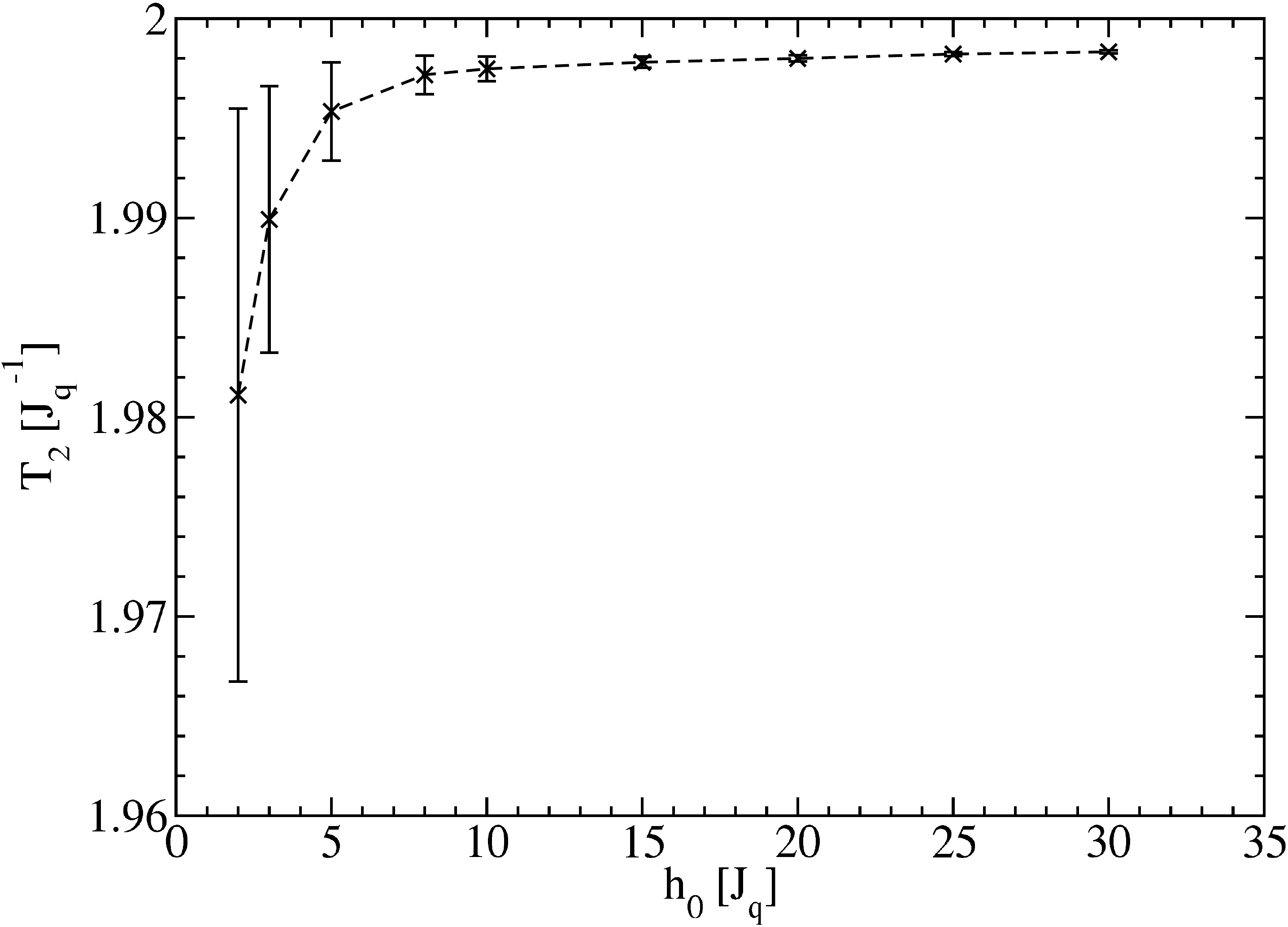}
  \caption{Spin-spin relaxation (or dephasing) time $T_2$ as a function of the
    external magnetic field $h_0$. The values for $T_2$ are obtained by
    fitting the function defined in Eq.~\eqref{eq:03_SxSx_fit} to the DMRG
    autocorrelation $\braket{S^x_0(t)S^x_0(0)}$ for $N=499$ bath spins. The
    error bars represent the errors of the fit. Note the small scale of the
    $T_2$-axis.}
  \label{fig:9}
\end{figure}

The so-called spin-spin relaxation time $T_2$ quantifies the transversal
dephasing. We determine it from a fit of the function
\begin{align}
  \braket{S^x_0(t)S^x_0(0)}&= \frac{1}{4}\cos\left(\omega t\right)
  \e^{-\frac{t^2}{2T_2^2}}
  \label{eq:03_SxSx_fit}
\end{align}
to the DMRG data or to the classical data. We refrain from a detailed analysis
for intermediate or weak fields so that both the quantum and the classical
approach serve the purpose equally well. In the following, we use the DMRG
results for $N=499$ bath spins as input. The extracted dephasing times $T_2$
are plotted in Fig.~\ref{fig:9} up to very large values of $h_0$. The value
$h_0=2\,\Jq$ is the lowest magnetic field for which a fit of the DMRG data to
Eq.~\eqref{eq:03_SxSx_fit} works reasonably well.

As indicated by the small error bars, the function defined in
Eq.~\eqref{eq:03_SxSx_fit} approximates the autocorrelation function
$\braket{S^x_0(t)S^x_0(0)}$ for $h_0>5\,\Jq$ extremely well. The error of the
fit increases for smaller values of $h_0$ where Eq.~\eqref{eq:03_SxSx_fit}
starts to deviate from the temporal behavior of the autocorrelation
$\braket{S^x_0(t)S^x_0(0)}$. However, with respect to the very small scale of the
$T_2$-axis in Fig.~\ref{fig:9}, we conclude that $T_2$ hardly depends on
$h_0$. This finding establishes that the dephasing time $T_2$ is solely
determined by the intrinsic time scale $\Jqt$ of the hyperfine
interaction. The external field $h_0$ does not have any influence on the
dephasing time $T_2$ once it is sufficiently large, $h_0\gtrapprox 2\Jqt$.
This finding agrees perfectly with the line width and shape of the
autocorrelation spectrum of the central spin, which was found to be essentially
independent on a sufficiently large magnetic field for large magnetic field
\cite{Hackma2014}
      
Many further studies of related Hamiltonians are called for at this stage. For
instance, the investigation of the dependence of $T_2$ on various
distributions of couplings suggests itself. Similarly, various extensions by
additional couplings such as quadrupolar interactions for larger spins
\cite{sinit12} or dipole-dipole couplings \cite{Merkul2002} between the bath
spins will help to understand the spin dynamics in quantum dots
quantitatively.

\section{Conclusion}
\label{sec:conclusion}

The goal of the present paper was twofold. First, we aimed at establishing
efficient methods to compute the dynamics of the central spin in the central-spin model (CSM). Second, on the conceptual level, we provided strong evidence
that the correct treatment of conserved quantities is important. This applies,
in particular, to the total spin and the total energy.

We studied a semiclassical, Langevin type approach to the central-spin
dynamics and a classical simulation for it. To verify the agreement of these
descriptions with the full quantum-mechanical ones, we compared their results to
the quantum results obtained from the previously introduced DMRG
approach.\cite{fried06,Stanek2013}

For the simple semiclassical model (Langevin 1), it turned out that the
results deviate rather quickly from the quantum CSM. The explicit
incorporation of conserved quantities such as the total spin improved the
agreement of the semiclassical model (Langevin 2 or 3) with the quantum result
substantially for zero magnetic field. Still, this approach does not conserve
the total energy because no backaction effects from the central spin on the
bath are incorporated. Additionally, on the technical level, it is a major
caveat that the bath fluctuations have to be known from an additional,
external source.
 
Neither restriction holds for the fully classical simulation based on the
solution of the classical equations of motion for both, the central spin and
the bath spins. The classical model shares the same conserved quantities with
the quantum CSM.\cite{Gaudin1983,Chen2007} Clearly, no further input from
other sources is required. We found that the agreement of the classical ansatz
with the quantum results depends on the strength of the external field applied
to the central spin. We could not establish classical behavior for large $N$
\emph{independent} of the considered time $t$ and of further parameters such
as the magnetic field as suggested by a spin path integral
argument.\cite{Chen2007} Without external field or for small fields, quantum
fluctuations induce a quantitative deviation between the classical and the
quantum model. For large fields, i.e., about twice the size of the
root-mean-square of the Overhauser field, the classical simulations agree very
well with the quantum results. We emphasize that in this regime the DMRG also
works extremely well and can access long times.

By either the DMRG or the classical calculation we determined the relaxation
rates $1/T_1$ and $1/T_2$ in the regime of strong fields. The spin-lattice
relaxation rate is essentially zero while the spin-spin relaxation rate is
given by $\approx\Jq$ independent of the magnetic field as long as it is
strong enough.

Future application of the semiclassical, the classical approach or the full
quantum-mechanical DMRG to the central-spin model comprise the simulation of
coherent control pulses and trains of such pulses which extend the dephasing
time of the central spin. Furthermore, the model can be extended by passing to
larger spins, adding anisotropic couplings such as quadrupolar terms, or by
including dipole-dipole interactions between the bath spins. In this way, the
understanding of the dynamics of electron spins in quantum dots can be put on
a more and more quantitative basis.
 
\acknowledgments

We thank F.\,B.~Anders, A.~Faribault, J.~Hackmann, and J.~Stolze for many
useful discussions. The authors are indebted to J. Hackmann for provision of
the Chebyshev expansion results. Financial support of the Studienstiftung des
deutschen Volkes (DS) and the Deutsche Forschungsgemeinschaft in project UH
90/9-1 (GSU) is gratefully acknowledged.

\end{document}